# Critical analysis of data concerning *Saccharomyces cerevisiae* free-cell proliferations and fermentations assisted by magnetic and electromagnetic fields

Jordan Hristov[1], Victor.H.Perez[2]

**Abstract** – The review analyses studies on magnetically assisted proliferations and batch fermentations with *Saccharomyces cerevisiae* yeasts. The results available in the literature are contradictory and show two tendencies: magnetic field suppression of the cell growth and positive effects in batch fermentation with increasing both biomass and metabolite production. The amount of data analyzed allows several concepts existing in the literature to be outlined and critically commented. Further, a new concept of magnetically induced micro-dynamos, recently conceived, is developed towards a unified explanation of the results provided by proliferation and batch fermentation experiments.

## I. Introduction

*I.1 Magnetically assisted free-cell process:
The basic idea*

Magnetically assisted bioprocesses are quite attractive for boosting classical technologies toward better performance in production of food and fuel by simple external magnetic field generation [1], [2]. The free-cell processes form special branch in this direction because they have to combine in almost mechanistic way the existing technological achievements in fermentation and magnetic field generation. The latter especially means that no special reactors and magnetic systems have been designed in contrast to the processes employing magnetic bio-carriers for immobilization of living cultures [2], [3]. In general, there are principle trends in magnetically activated free-cell processes [2]:

    a) Proliferations or fermentations in vessels entirely surrounded by magnetic fields and,

    b) Batch fermentations with external loop for periodic magnetization of the culture medium.

The basic idea in the magnetically assisted bioprocesses is a non-conventional stimulation of classical processes producing biomass, ethanol, biodiesel [4], etc. The main expectations are that field action (assistance) may force or decelerate the growth and metabolic production of the living organisms. *Saccharomyces cerevisiae* and *Spirulina platensis*, for examples, related to the mass production of food and fuel. The analysis performed in this review focus on some crucial questions, among them

- Is it possible to stimulate the *Saccharomyces cerevisiae* growth and metabolitic activity with external magnetic or electromagnetic fields?
- How, the fermentor has to be designed and what are the basic conditions that has to be satisfied for correct process performance?
- How the field effects, decelerating or boosting, can be explained?

Following Hunt et al. [1], the fields acting on a certain volume with living cultures at issues can be classified as:

1) **Magnetic fields** (the in the *Near-field* regime) created by conventional coils (solenoids) and operating with steady-state (no time-varying), slowly varying (low-frequency range, commonly termed as **extremely-low frequency range**) in time and pulsing modes. The last one is mainly applied in experiments with extremely high (from a laboratory level standpoint) steady-state magnetic fields.

2) **Electromagnetic fields** with both electric and magnetic components, with ratio between $0.1$ and $1.0$, operating in the *Far-field* regimes with typical oscillating frequency of $100\,kHz$ or more.

EMF frequencies in the cells are normally in the range of extremely low frequencies (ELF)-see Table I for definitions. Static magnetic fields (MF) can also to be considered as specific ELF because if the liquid based cell cultures are moving (mixing in fermentation experiments) then refereeing to the Faraday law, there are induced eddy currents of frequencies depending on the relative motion between the conducting medium and field lines.

Compiling and analyzing data from various literature sources is not easy due to many reasons, among them:

- incomplete information about the experimental equipment used,
- insufficient information taken during the experiments,

- hard to create duplications of experiments,
- Sometimes missing direct information regarding cell proliferation, metabolic activity, physical properties of the culture medium, etc.
- It is extremely difficult to determine which responses directly result from EMF and which are due to incorrect experimental conditions

TABLE I.
MAIN EXPERIMENTS ON ALTERNATING FIELD ASSISTED *S. CEREVISIAE* PROLIFERATION. FREQUENCY RANGES OF FIELD ACTING OF LIVING CELLS.
After Funk et al.[5]

| Common term | Frequency range | Source of radiation |
| --- | --- | --- |
| Extremely Low (**EL**) | $0 - 300 Hz$ | Domestic mains supply Laboratory units Experimental devices for cell proliferation and fermentations |
| Low, Middle and High | $30 kHz - 30 MHz$ | Communications by portable radio stations. Industry control systems |
| Very, Ultra, High | $30 - 300 MHz$ | Radio and TV broadcasting |
| Super High | $300 MHz - 30 GHz$ | Cell phones and Satellite communications |

The literature dealing with magnetic field effects on biological systems is full of a mystifying collection of explanations, models, and incomplete explanations. In this review, we concentrate mainly on *Saccharomyces* yeasts findings and try to find common links and explanations of the mechanisms by which MF or ELF affect certain biological system, i.e. a proliferation experiment or a fermentation run. The literature sources, for instance, refer to numerous ''frequency windows'' without any account for the thermal effects of radiation by the specific absorption rate (SAR) of yeasts themselves or the specific culture medium in particular. However, in the case with ELF the energetic threshold to produce cell-specific information at defined frequencies can be much lower than that required for unspecific heating of the living organisms [5].

The most complex, and currently the most speculative, component of the studies on magnetic field effects on biological systems are the so-called non-thermal effect addressing intracellular action, protein folding, ion channel activation, etc. However, the papers reported up to now do not clarify completely the effects of MF on living organisms and bypass unresolved problems and avoid discussions on different results, contradictory in some cases.

## II. *Saccharomyces cerevisiae*

It is basically the most useful yeast since ancient times in bread and alcohol production. It is one of the most intensively studied eukaryotic organism in cell biology comparing to *Escherichia coli* as model prokaryote. The cells are round to ovoid, 5-10 µm in diameter and reproduce themselves by a division process known as **_budding_**. *S. cerevisiae* can grow on glucose, maltose and trehalose and cannot develop on lactose and cellobiose. The ability of yeasts to use different sugars can differ depending on whether they are grown aerobically or anaerobically: it was shown that galactose and fructose were two of the best fermenting sugars. Yeast cells do not need regulated $CO_2$ atmosphere and they usually grow between 25 and 30°C. Three typical pictures taken by electronic microcopy are shown in Fig. 1

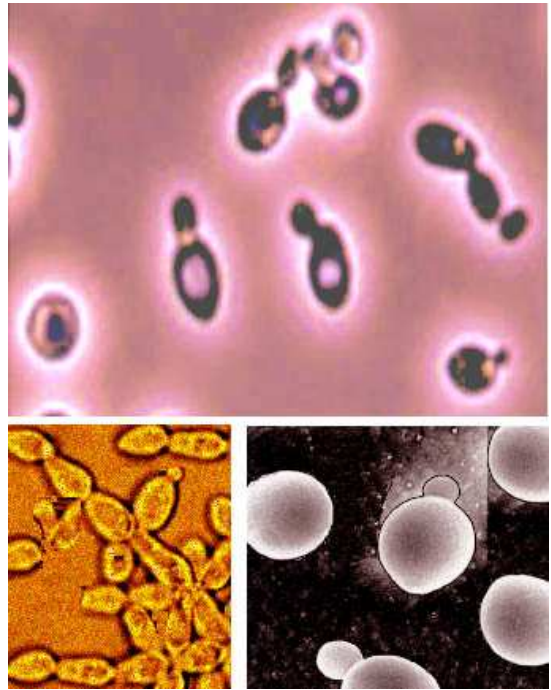

Fig.1. Samples eletronic microscope photos of *S.Cerevisiae* yeasts. Not to scale. By courtesy of Prof. Viara Ivanova, Bulgaria

*Saccharomyces cerevisiae* is a good eukaryotic model organism for various biological studies. It is quite suitable in research because of its greater amount of genetic material than bacteria, rapid growth, well defined genetic system, and easy handling. Besides, the regulation of the cell cycle seems to be similar in all eukaryotic cells, which makes yeast cells suitable for studying the linkage between DNA damage and the cell cycle [6], [7].

These are common facts, but now we will stress the attention on other properties that might help us elucidating the magnetic field coupling in cell growth. Especially we refer to characteristics of suspensions containing yeasts allowing understanding how these media transport either electric or magnetic fields. In this context we refer to work of Woodward and Kell [8] on the dielectric properties of *S. cerevisiae*. According to these authors, when a field of appropriately low frequency is applied to a suspension of cells contained between two or more macroscopic electrodes, the charging of the membrane capacitance may also cause a lesser but effective "amplification" of the macroscopic field across the membrane (see further the comments in this work). Further, they demonstrated that in the presence of a very modest exciting field of only $\pm 1 - 2.5 V$ at $15 - 20 Hz$, sizeable 3rd and 7th harmonics could be observed under conditions in which the linear impedance was (as expected) voltage-independent. Moreover, Tatabe et al. [9] reported optimums frequency range of $15 - 30 Hz$ where the depletion of the glucose in the culture medium and disappearing of odd harmonics are strongly interrelated. These results allow by monitoring the nonlinear dielectric properties the status of the cells (rest or active) to detected and monitored.

## III. BASIC KNOWLEDGE ABOUT MAGNETIC FIELD EFFECTS ON CELLS

We will briefly outline some basic information concerning physical and biological aspects of the magnetic field coupling to cell growth and biology. This approach allows creating a background for further comments of the results concerning especially the *Saccharomyces cerevisiae* proliferations and fermentations.

### III.1. Basic effects of DC EF and MF on cells and aqueous solutions

### III.1.1. Where EF and MF act?

Because of the high conductivity of liquids of the culture media the exposure to electric fields (EF) generated around either by electric field generator or induced by eddy currents (due to relative motion of the cultures and static magnetic fields or by ELF). Because EF do not penetrate the bodies (cells), they create surface charges. i.e. the cells are shielded. Considering the dimensions of a cell, then the thickness of a cell membrane (10 nm), then with a 0.1V difference we get a field strength of about 106–107 V/m.

The magnetic fields, in contrast to the **EF**, however, can penetrate the membrane and go deeper into the cell, i.e. affects the whole body. The EF for electroporation of drugs or genetic material should be in the range of $1000 - 5000 V/cm$ [10]. Despite these high-field strengths required to penetrate cell membranes, the other important magnetic effect is the ability of internal signaling pathways through sensors mechanisms of the membrane.

Because the magnetic field stimulations of bioprocesses use "technical fields" the natural MF (i.e., that of the earth equivalent to about $5 \times 10^{-5} T$) will be skipped in this analysis and they will be analyzed further in this review.

### III.1.2. Coupling MF effect into cell growth

According to Funk et al. [5] there are principle approaches (sides of discussion considering specific effects) the in analyses of coupling magnetic fields and cell biology: physical approach and biological point of view, briefly commented next.

### III.1.2.1. Physical approach

This point of view based on fundamental knowledge of the solid state physics and classical electrodynamics. This approach mainly focuses on mechanisms by which MF affect living systems in the context of Faraday's Law: changing magnetic field is associated with a changing electric field, and *vice versa*. This approach needs at least flux densities in the $mT - T$ range to explain the effects.

- **Faraday coupling**. Beyond, the Faradays Law, the main opinion about the magnetic effects behind magnetic address direct mechanical action on the molecules, such as : i) transient rotational motions of substances (due to magnetic moments of rotation, i.e. torques) in uniform magnetic field related to achievement of minimum energy states; ii) translations of paramagnetic or ferromagnetic substances due to magnetic gradients. Biological tissues are diamagnetic with few exceptions, thus their magnetic susceptibilities are close to vacuum, so these direct mechanical actions can be excluded. However, in accordance with the Faraday's Law the relative motion of a conducting medium and static magnetic, as well as static conducting medium and time-varying MF, yield electric currents and EF. Hence, any mechanical actions could be related to those created by surface charges on the cell membranes rather than to moments of rotation and field gradients. In this context, the induced electric currents by MF (Faraday coupling) as active agent are discussed by Schimmelpfeng and Dertinger [11]. Their experiments concerning proliferation of SV40-3T3 mouse fibroblasts and human HL-60 promyelocytes revealed that exposure to a sinusoidal $2mT$ ($50Hz$) magnetic field the cell growth was only affected by an induced electric field beyond a threshold between $4$ and $8 mV/m$ at $2mT$.

Because the magnetic component of EMF and static MF can penetrate deep into the cell, Faraday's induction law is also applicable, as demonstrated by reorganization of the electrostatically negative charged actin filaments. In this context, Cho et al. [12] have demonstrated that either at $1$ or $10 Hz$ field (i.e. in the ELF range) the microfilament structure can be disrupted: from an aligned form to separate globules. Increase in the frequency ($20 - 120 Hz$ do not affect the microfilament structure, that lead to the hypothesis that at low frequency the filament structure is close its mechanical resonance conditions (concerning, moment of inertia, viscosity of the surrounding liquid, etc). Beyond the resonance frequency, the increase in frequency of the forcing field, for instance, does not affect the structure, a fact well-known from the classical mechanics.

- **Larmor precession** (discussed further in this work) is one of the reasonable effects affecting to influence biological systems by MF, mainly those in motion like cell organelles and macromolecules. Precisely, since electrons are in motion in an orbit and have spin, the torque exerted produces a change in angular momentum perpendicular to that angular momentum, causing the magnetic moment to precess around the direction of the magnetic field rather than settle down in the direction of the magnetic field: this is the **Larmor precession.**

*III.1.2.2. Biological point of view*

With enormous number of charged particles and surfaces from single molecules to large macromolecules and cell organelles, the common explanation [13] of biological coupling of MF is effect of the ion transport through the ion channels of the cell membranes. However, we should take into account that the ions are neither in vacuum nor freely moving in liquids, but ions moving in a body. Hence, the commonly recognized effect of MF on the ion's Brownian motion and triggering of the ion-channel transport could be considered as an attempt to explain the magnetic coupling but not as definitive explanation.

On the other hand, the cell membrane is considered to be the key object for MF through effects on the rate of ion or ligand binding. Hence, the MF affects the receptor side affects and consequently the signaling chain involving $Ca_2^+$ transport and growth factors [14]. The biochemical reactivity of ions can be attributed to changes in the spatial orientation of movement or by changing Larmor precession frequencies [15].

*III.1.2.3. MF effects on aqueous solutions: physical and chemical aspects*

It is well-known, but still non-understood, that the magnetic field effect on both the physical and chemical properties of aqueous solutions of different salts and crystal growth rates of diamagnetic inorganic salts [16], [17], [18]. In general, water is *diamagnetic*. Besides, magnetic treatment is commonly used to a de-scaling treatment of water pipes even though the nature of this phenomenon is still unknown and matter of arguments [19]. These effects often persist up to several days after removing the applied field. In this context, it may suggest that those changes in the aqueous solution itself may promote biological effects.

*III.1.2.4. Static MF effects on cells*

Moderate-intensity static MF (non-time varying) may act on biological systems through re-orientation of the diamagnetic membrane phospholipids as it was suggested by Rosen [20]. This re-orientation results in deformed ion channels, thus affecting their activation kinetics. The studies of calcium channels support for this hypothesis. Moreover, there is a temperature dependency that is understandable on the basis of the membrane thermoropic phase transition. Additionally, that sodium channels are also affected by static MF to some extent.

*III.1.3. Other magnetic field effects*

- ***Magnetic effects on chemical reactions involving free radicals***. Free radicals are generated as intermediates in metabolism and may attack lipids, proteins and DNA. Hence, any rise in the free radical production could increase the chemical damage. Magnetic field of more than $1mT$ can have considerable effect on the kinetics and yield of chemical reactions using geminate radical pairs. Moreover, magnetic fields through their effect on the spin precession rates of unpaired electrons and consequent effects of the lifetime of radicals [21]. The magnetic fields can increase or decrease the precession rates between singlet an triplet spin-correlated states [22].

- ***Thermal effects.*** The extent of the thermogenic effects depend mainly on the magnetic field intensity which is associated with the specific absorption rate (SAR). Increase in temperature or thermal effects can result in changes in cellular functions, and, under extreme conditions can result in cell damage.

- ***Non-thermal effects***. There are numerous published papers that show different modifications of biological systems after exposure to weak EMFs [23]. There are several non-thermal parameters of EMF which are very important in analyses of field effects on cell. A few examples of these parameters [23] include frequency of the carrier wave, modulation frequency, near field-far field effect, polarization, duration of the exposure, continuous or pulse wave exposure, shape of the pulse and, finally, presence or absence of a static magnetic or electric field.

- *Electroconformational coupling* (ECC) for cellular enzymatic systems [23],[24] suggesting that the enzymatic systems, particularly those in membrane structures, receive, process and transmit high and medium level intensity periodic electric potentials. This approach considers four-state enzyme systems converting electric field energy into chemical potential energy under certain circumstances: the frequency and the strength of the applied field properly match the characteristics of the system. The ECC theory suggests that there are optimal frequencies and amplitude intervals of oscillating electric fields and optimal ligand concentrations for efficient coupling of the electric field to enzyme systems.

Optimal frequencies depended on the system under investigation and varied from $10 Hz$ to $1 MHz$ [24]. The externally applied electric fields with frequencies less than 1 MHz, can be amplified by cellular membranes [25]. Because the membrane is an insulator with respect to the external electric field of intensity $E_{ex}$, then the difference in the electric tension (voltage) across a membrane of a cell of radius $R$ is $\Delta U_m \approx 1.5 R E_{ex}$. Further, the electric field intensity in the membrane $E_m$ is the gradient of the voltage across the membrane of thickness $d_m$, i.e. $E_m = \Delta U_m / d_m = 1.5 E_{ex} (R/d_m)$. Hence, the intensity of the external electric field is amplified by a factor of $1.5(R/d_m)$ [23]. Computational models [26], for instance, suggest that under amplitude-modulated microwave electric field, the non-thermal effects are exhibited as induced low-frequency ion currents in the cellular sodium channels.

### IV. Magnetic Field Assisted *S. cerevisiae* Proliferation Studies

*IV.1. General idea and trends*

The main idea in magnetically assisted processes involving *S. cerevisiae* is to boost the process of cell growth or metabolic activity thus achieving high yields. Many articles were written on the problem but we will comment results of some of them where information and analyses are of primary importance for elucidation of the problem [27], [28], [20]. Table II summarizes most of the data elucidating the experiments performed up to the end of 2010. The further analysis follows the scheme:
  i) *S. cerevisiae* cell proliferation in ELF of industrial frequency
  ii) *S. cerevisiae* cell proliferation in static (or single pulse) MF of low and high flux density
  iii) Fermentations in reactors neither completely encircled by magnetic fields or by recirculation.

TABLE II.
MAIN EXPERIMENTS PERFORMED ON ALTERNATING FIELD (ELF) ASSISTED *S. cerevisiae* Proliferations

| Strain | Field | Duration | Ref. |
|---|---|---|---|
| *S. cerevisiae* Not specified | Helmholtz pair ($5 cm$ ID) 10% non-uniformity around the central coil axis). $1-100 Hz$ with amplitude $0.5-0.9 mT$ | $10 h$ | Mehedintu and Berg [27] |
| *S. cerevisiae* | Helmholtz pair $50 Hz$, $0.28-12 mT$ | $9 h$ | Fiedler et al. [28] |
| Numerous strains See the original article for details | Square flat poles ($100 \times 100 mm$) and gap of $20 mm$ $60 Hz$, $0.6 T$ Petri dishes ($80$ and $50 mm$) | 3 days growth | Shimizu et al. [37] |
| Prototrophic tetraploid strain (a/A) S. cerevisiae CCY 21-4-59 | $50 Hz$ ($5.2; 7.9; 10 mT$) | 24 min | Novak et al. [30] |
| Haploid yeast *S. cerevisiae* (wild type) and three derivative strains mutated for the genes HDF1, RAD52 and their double combination | Helmholtz pair, $400 mm ID$ $50 Hz$ ($0.35$ and $2.45 mT$) Axial orientation (i.e. parallel to the axis of the tube containing the culture medium) | $96 h$ | Ruiz-Gomez et al.[31], [32] |
| *S. cerevisiae* beaker's yeast (Pak Food Produced Com.). | Helmholtz coils ($20 cm ID$ $15 Hz$ pulsed field ($1.1 mT$) at the center of the coils | $29 h$ at $100 r.p.m$ of the shaker | Gulbandilar [33] |

*IV.2. Alternating magnetic fields with industrial frequencies ($50-60 Hz$)*

Mehedintu and Berg [27] reported different yeast growth rates depending on the magnetic field amplitude. The cell number decreased at $0.2 mT$ is applied and increased of about $25\%$ when the amplitude is $0.5 mT$. Moreover, two positive windows were detected (at $15 Hz$ and $50 Hz$) while in this range only inhibition was observed at $40 Hz$. The positive windows correspond to about $20\%$ increase in the cell number for $10 h$ continuous exposure to the field. However, the exposure to the field resulted in decrease in both the volume and the radius of distribution of the stimulated

*J.Hristov ,V.H.Perez*

cells. Further, the heat effect is quite important because the increase in the temperature may augment the field effect at low frequency (at $15 Hz$) with about $25\%$, while the positive effect at $50 Hz$ disappeared. In general, the study of Mehedintu and Berg [27] is undeveloped with many unclear moments and obscure experimental conditions even though this is quite frequently cited article. To this end, the temperature effect is easily to be explained if the physics of the magnetic field and the culture medium is well understood. However, any deep analysis is missing. We will try to explain this from the point of view of the micro-dynamo concept (MDC) conceived recently [2] concerning Faraday couplings in the conducting liquid phase only (see the discussion section of the present work) and excluding the so-called *non-thermal effects*.

Similar experiments have performed by Fiedler et al. [28] in the lower part of the frequency range used in [27]. The outcome was about $17.6\%$ increase (with respect to the control culture) in the biomass with $0.5 mT$ magnetic field exposure that was detected as the only positive window. An important fact supporting the micro-dynamo concept, and related to the experiments of Mehedintu and Berg [27] is that the cell proliferation can be enhanced at fields lower than $1 mT$.

The negative effect of the Berg's group at $50 Hz$ on the living culture with increase in the field amplitude was confirmed by the experiments of Novak et al. [30] and, in general, by those of Fojt et al., [35] with Paracoccus *denitrificans* [35], [36] with *E.coli* [35],[36] under the same conditions [2].

Novak et al. [30] found that MF decreases the number of yeast, and slowed down their growth after exposures to $5.2$, $7.9$ and $10 mT$ at $50 Hz$ for $60 \min$. These authors used the prototrophic tetraploid strain (a/A) S. cerevisiae CCY 21-4-59. They suggested that it seems that the MF killed a part of yeasts and the bigger part of them survives and continues in their growth. Some results of Novak et al. [30] are shown in Fig. 2.

In addition, no alterations in the growth of *S. cerevisiae* were found after exposure to sinusoidal $50 Hz$ ($0.35$ and $2.45 mT$) [31], [32]. These authors [32] reported that the extremely-low frequency MF (the same conditions as in [31]) induces alterations in the growth and survival of *S. cerevisiae* strains deficient in DNA strand breaks repair (hdf1, rad52 and rad52 hdf1). However, the MF treatment does not induce alterations in the cell cycle and does not cause DNA damage. Some results shown in Figure 3 indicate that the growth curves for the wild type (wt) yeast strains under continuous MF ($2.45 mT$, $50 Hz$ MF) are in all case above those of control (non-exposed) ones. In this context, Luceri et al. [36] reported that extremely-low frequency MF ($1$, $10$ and $10 mT$, $50 Hz$, $18 h$) do not induce DNA damage or affect gene expression in *S. cerevisiae*. These results confirm the report of Shimizu et al. [37], performed in $60 Hz$ high density field (see Table II) on mutagenicity and growth of yeasts.

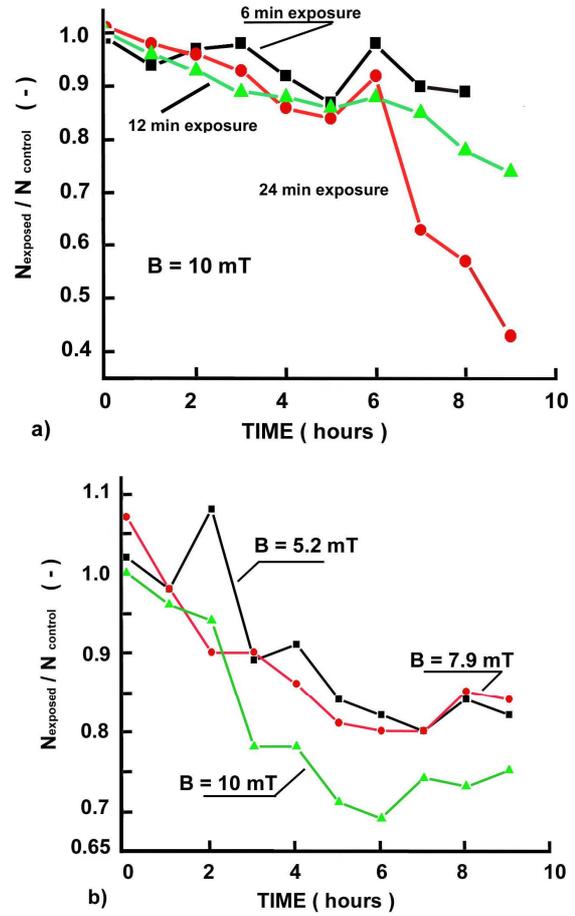

Fig.2 Results of Novak et al. [30]. Graphical representation (present authors) of the data collected in Table 2 and Table 3 of the original work.

It is noteworthy to comment not well-known work of Gulbandilar [33] where the S-shaped growth curves (see Fig.4) were modelled by the Verhulst's logistical curve [38] ( a generalization of the Malthusian Dynamics ), namely

$$\frac{dN}{dt} = \mu N , \quad N = N_0 e^{\mu t} , \quad N = N_0 \text{ at } t = 0 \quad (1a,b)$$

applied to the exponentially growing phase. The simplest approximation is in the form

$$\frac{dN}{dt} = \mu N - r N^2 \quad (2)$$

That yields

$$N = \frac{\mu}{r + \left(\frac{\mu}{N_0} - r\right) e^{-\mu t}} \Rightarrow N = \frac{N_s}{1 + k e^{-\mu t}} , \quad (3a)$$

$$k = \frac{N_s - N_0}{N_0} \quad (3b)$$

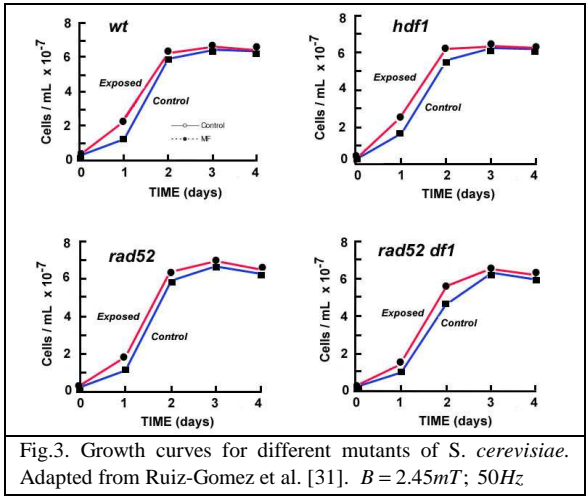

Fig.3. Growth curves for different mutants of *S. cerevisiae*. Adapted from Ruiz-Gomez et al. [31]. $B = 2.45 mT$; $50 Hz$

The plots in Fig. 4 appear to have different characteristics after 11-12 hours of incubation period that allowed Gulbandilar to suggest that the cell division gets into a different process. In general, it was found out that the pulse magnetic field applied reduces the growth only in the exponential phase and there is shift in the time scale of about $1.25 h$ between the exposed and the control growth curves. The results of the control growth curves were fitted as $N_a = 10\left(1+33.5e^{-0.78t}\right)^{-1}$, while those for the magnetically boosted as $N_b = 22.9\left(1+20.2e^{-0.51t}\right)^{-1}$. The growth was modelled as a sum of

$$N = \frac{10}{1+33.5e^{-0.78t}} + \frac{22.9}{1+20.2e^{-0.051t}} \quad (4)$$

The developed equations could serve as examples for further fitting of experimental results. In this context, no modelling of magnetically exposed growth concerning *S. cerevisiae* was reported except the approach of Otabe [39] based on the Neumann's idea [40] discussed further in this work.

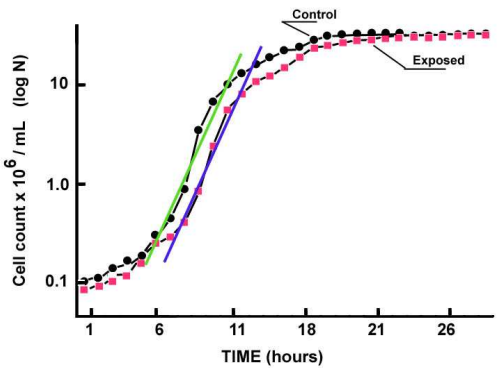

Fig. 4. Growth of S. cerevisiae in a pulsed filed $15 Hz$. Adapted from [33].

### IV.3. Radio–frequency (RF) and extremely high frequency (EHF) experiments

The radio-frequency (electromagnetic fields) EMFs are more interested as sources of environmental electromagnetic contaminations rather than as tools oriented to boost biotechnological processes. However, to fill more facets in the presentation of *S.cerevisiae* exposed to various non-ionizing radiations, we refer to increase in free radical production as a yeast response [41]. Samples were loaded in glasses micropipettes of $20 mL$ placed in the cavity (pulsed cavity DLQ, Bruker, France) of the ESR system at $22.8^o C$ with a $1 kW$ travelling wave-tube amplifier (TWT) as exposure device. Typical frequency was $9.71 GHz$, with a pulse length of $9000 ns$ and a duty cycle of 1% and exposure times of $20 min$. These experiments performed with water-soluble spin trap a-(4-pyridyl-1-oxide)-N-t-butylnitrone and the lipid-soluble N-tert-Butyl-$\alpha$-phenylnitrone showed an increase of spin adduct production both in low power density exposure ($SAR < 4 W/kg$) and in thermal conditions ($SAR > 4 W/kg$). The overall results suggest an increase of the free radical production in the intra cellular compartment but no effect on the yeast vitality was found.

On the other hand, in the experiments of Gos et al. [42] yeast *Saccharomyces cerevisiae* (strain WDHY376) were exposed to electromagnetic fields in the frequency range from $41.682 GHz$ to $41.710 GHz$ at low power densities ($0.5 \mu W/cm^2$ and $50 \mu W/cm^2$) to look for non-thermal effects on the division process. The exposure chamber was a copy of the design of Grundler [43]. The yeast was exposed on agar surfaces, and division was recorded via time-lapse photography during the exponentially growing phase. The data from several independent series of exposures and controls reveal no consistently significant differences between exposed and unexposed cells.

Recently, Vhrovac et al. [44] reported experiments on $905 MHz$ microwave (MW) radiation (similar to that of cell phones) on colony growth of the yeast *Saccharomyces cerevisiae* of different strains and exposure times ($15$, $30$ and $60 min$). The experiments reveal that the wild-type strain FF18733 did not show statistically significant changes in colony growth compared to the control sample. However, irradiated strains FF1481 and D7 demonstrated statistically significant reduction of colony growth compared to the control irrespective of the exposure time. Moreover, it was reported that $60 min$ exposures reduce the colony growths of strains FF18733, FF1481 and D7 of about $19.30 \pm 2.06\%$, $56.37 \pm 1.49\%$ and $34.29 \pm 3.21\%$, respectively. The author suggested that microwave radiation at low SAR level led to DNA damage macroscopically performed as reduced growth rates. Moreover, they comment unpublished preliminary experiments with modulated RF/MW radiation with a low SAR value did not affect either the rate of gene conversion nor reverse mutations in strain. In this context, they refer to [45] where $900\, MHz$ MW radiation was used without any mutations or recombinations and genotoxic effects, as well.

The millimeter wave range ($53.8\, GHz$ and power flux density $1\, mW/cm^2 \div 10\, mW/cm^2$) was investigated by Usatti et al. [46] on the multiplication and biosynthetic activity of oleogene components of yeast strain *Saccharomyces carlsbergensis CNMN-Y-15* (as active producer of sterols). In general, the effects depend on many operating conditions such as: mode of irradiation, duration of exposure, growth phase. The proliferation run (in $1L$ Erlenmeyer flasks with $0.2L$ YPG substrate medium) lasted $72\, h$ at $28^o C$ with different MW exposures ($1 \div 60\, min$) in both continuous and pulsed modes of irradiation. The results reveal yeast stimulation with about $30 - 59.4\%$ when the exposure times are from 1 to $10\, min$ and periodical irradiation mode where the most significant effects were observed in the first $24h$. In contrast, the continuous irradiation inhibits the yeast multiplication in the first $48h$

### IV.4. Proliferations in low and moderate intensity static fields

Van Nostran et al. [47] have reported the inhibitory effect of static magnetic field at different osmotic pressures (see experimental details in Table III). These authors have found that increasing the osmotic pressure produced the inhibitory effect of the high magnetic field is masked by the osmotic pressure at the higher temperature (see Fig. 5). They suggested that that decreasing the water in the cells allows multiplication to proceed readily at the high temperatures.

*Saccharomyces cerevisiae* biomass growth was investigated by Muniz et al.[48] in fields created by permanent magnets and conditions simulation of ethanolic fermentation This report claims that for $24h$ exposure to a field ($550\, mT$) of $200\, ml$ culture medium the biomass growths in both the exposed and the control samples were almost the same during the first $8h$ of fermentation. The growth of the magnetized yeast was high while than of the control (no-magnetized) sample was moderate. In fact, the magnetized yeast demonstrated $3.7$ times superior growth rate. At the same time, the biomass value for $24h$ exposure attained a level of $2.5$ times greater than that of the control. Moreover, the sugar depletion in the magnetized yeast culture was higher that in the control one beyond the point of $8h$ to the end of the process. In this context, the depletion rate of the average glucose level in the magnetized culture attained a level of 32.2% greater that than in the no-magnetized culture.

TABLE III.
MAIN EXPERIMENTS PERFORMED ON STATIC FIELD ASSISTED *S. cerevisiae* Proliferation

| Strain | Field | Experiment duration | Ref. |
|---|---|---|---|
| Wild *S. cerevisiae* strains (see the text) | Parallel plate poles $B = 4600\, Gauss$ | $24;\, 48;\, 72h$ | Van Nostran et al. [[47] |
| Numerous strains See the original article for details | Square flat poles ($100 \times 100\, mm$) and gap of $20\, mm$, $0.5T$ Petri dishes ($80$ and $50\, mm$) | 3 days growth | Shimizu et al. [37] |

As a portion of these experiments [49] it was performed a test at $110\, mT$, $220\, mT$ and $350\, mT$ field exposures under the same experimental conditions (vessel volume and speed of stirrers). The results can be summarized as it follows:
a) The non-exposed cell cultures exhibit a growth greater than those of the exposed yeast.
b) The exposed at $110\, mT$ culture resulted in a lower biomass development than the non-exposed one for the first $12h$.
c) The second $12h$ of the inoculations revealed a magnetically induced growth. The cultures exposed to $220\, mT$ showed 3 peaks of the absorbance (spectrophotometric measurements at $610\, nm$ wavelength light), while at $110\, mT$ only 2 peaks appeared.
d) The produced $CO_2$ volume showed higher pressure values in the exposed samples that the control ones, after $8h$ inoculation: at $110\, mT$ the released gas was only of about $7\%$, while at $220\, mT$ the gas was $21.3\%$, more than the control volumes.

In general, Motta et al. [48], [49]explain the observed magnetic stimulation of the process on the basis of Blanchard and Blackman [50] interpretations attributing the holistic effect in the yeast performance to magnetic effects on certain proteins, more specifically to enzymes.

### IV.5. High static or single pulse magnetic fields

In contrast to the experiments with industrial frequency fields with low-intensity of the magnitude, no proliferation experiments were performed with static magnetic fields that can be easily generate under laboratory conditions by permanent magnets or coils. The efforts have been oriented to extremely high fields acting as pulses or over short periods of time on the yeasts. The experiments of Anton-Leberre at al. [29] indicate that no effect on the cellular processes in the *Saccharomyces cerevisiae* yeast can be observed: short-time exposures ($30\,min$) to $14T$ and a long-time exposure ($8h$) to $16T$ do not result in any effects on the yeasts. Similarly, yeast cultures exposed to a pulsed (sub-seconds) high intensity magnetic field ($37T$ and $55T$) as well as to $20T$ field pulses repeated one to four times do not changes their intracellular function and metabolic activity. The latter was confirmed by analysis of the genome-scale protein expression with both long-term exposure ($14T$ and $16T$) and pulsed magnetization ($37T$ and $55T$). Similar, results, i.e. no field effects, were detected through viability and viability and cell morphology tests by coloring cells by Methylene blue and exposure to 4 pulses of $20T$.

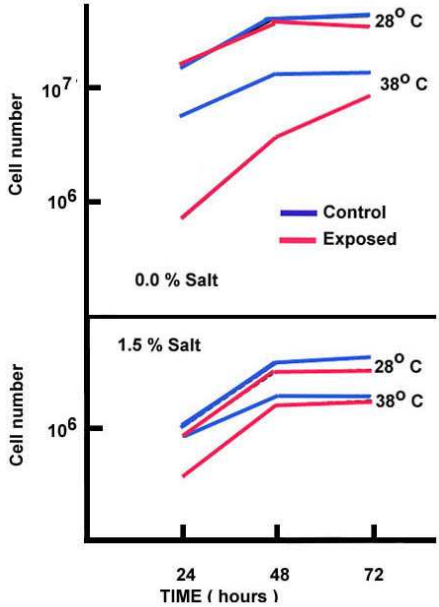

Fig. 5. Simultaneous effect of a magnetic field and temperature on *S. cerevisiae* cell growth. Adapted from [47]

In the above context, we have to mention the results of Iwasaka et al. [51] where the rate of yeast proliferation under the magnetic fields ($9-14T$) decreased after $16h$ of incubation in relation to unexposed control group. Moreover, it was showed [52] that inhomogeneous MF ($4.1-8.5T$) have selective effects on cell growth, cell cycle and gene expression in S. cerevisiae with deletions in transcription factors. They observed that MF increases growth only following the multicopy suppressor of SNF1 (sucrose non-fermenting 1) (Msn4) or splir finger protein 1 (Sfp1) deletion, associated with decreased G1 and G2/M and increased S phase of the cell cycle.

In this specific case (high static fields) we especially address the work of Otabe et al. [39] the activity of the yeast cells is suppressed by the strong fields and the glucose is not suffciently converted to the ethanol (resulting in sweet taste). For example, much slower fermentation could be achieved at low temperature and strong magnetic field. The time dependence of the number density of the yeast cells in the suspension (counted by standard plate count technique) was analyzed in terms of Malthus equation as given by

$$\rho_{yeast} = \rho_{0-yeast} \exp\left(k_{prof} t\right) \qquad (5)$$

where $\rho_{0-yeast}$ is the initial number density of the yeast cells at initial time $t=0$; $k_{prof}$ is the growth (proliferation) coefficient.

In general, the values $k_{prof}$ in magnetic fields are lower than those obtained for the control non-exposed volumes. It was observed that the differences in the growth coefficient became large as the field intensity was increased and become almost constant beyond fields of $5T$. The growth coefficient variation with field and temperature was expressed in an Arrhenius-like relationship [39], namely

$$k_{prof(B)} = k_{prof(B=0)} \exp\left(\frac{m^*}{k_B T} B\right) \qquad (6)$$

where $k_B$ is the Boltzmann constant and $m^*$ is the magnetic growth activation moment which represents the sensitivity to magnetic field [40]. Large values of $m^*$ mean that the growth activation is sensitive to the magnetic field. The work of Otabe et al. [39] is typical as a proliferation study, but gives an example how data have to be correlated with the intention of extraction of more logical and well organized data.

## V. Fermentation Processes

### V.1. Low and moderate intensity static fields

In contrast to the proliferation studies searching magnetic field effects at micro levels, the performed fermentation studies directly copied the classic fermentation process of ethanol fermentation [54] in vessels entirely exposed to external magnetic fields [48], [49], [54]. Two $200\,mL$ reactors (control vessel and magnetically exposed) performed simultaneously ethanolic fermentation. The magnetic field exposure for the stimulated fermentation was established at $550\,mT$. The results were taken from eight identical runs by continuous monitoring of ethanol and glucose concentration, and of the yeast growth. In general, the ethanol production and the glucose uptake are augmented in the magnetized culture by $30\%$ and $20\%$, respectively. The glucose consumption was almost complete at the end of the magnetically assisted fermentation run in contrast to the control (non-magnetic) counterpart. Further, the rates in the ethanol production and the yeast growth attained values of 3.4 and 3.0 times greater than those in the control fermentation runs. In this contest the ethanol yield per gram of glucose, per gram of biomass reached values of 4.2 times greater than those in the unexposed runs. These results are supported by the *glucose/biomass ratio* which was only 1.17 times higher than in the control fermentation.

Motta et al. [55] clearly indicated that the significant changes in the performance of the exposed fermentation were observed after an initial lag phase of $4h$ up to the end ($24h$) of the process. This supports the hypothesis of efficient field effect on the yeast growth and ethanol production in the exponential phase. The final ethanol average content was about 2.5 times higher than in the control runs; the biomass content was about 2.4 time higher.

The works of Motta et al. [48], [49], [54], [55] clearly denote that while in the unexposed fermentation the ethanol may alter the transport in the cell membranes and unfold the cytosolic globular enzymes [56] hindering the cell growth and proliferation, *this was not observed in the magnetized yeasts*. The boosting effect of the magnetic field exposure was not explained with other hypothesis of potential mechanism at micro or macro level. However, if we will try to look at the results from the position of the MLD concept developed in the Discussion.

Similar experiments of Galonja-Coghill et al. [57] show that fermentation aided by a static magnetic field (see Table IV for additional details) results in cell density of about $5.5 g/L$ (with final cell densities in control about $3.3 g/L$.) maximum ethanol concentration of $44 g/L$. The average productivity attained was about $2.75 g/L$ per hour, with about $71.1\%$ utilization of sugars. The lag phase in the ethanol production was about $4h$ with an initial concentration of $6 g/L$. For $4h$ exponential phase (i.e. $8h$ from the beginning) the ethanol production rated reached $19 g/L$, while after $12h$ the production was $39 g/L$. In general, it was reported that the ethanol production is not linear in time with $5, 12, 20$ and $23 g/L$ after $4, 8, 12$ and $16h$, respectively. The explanation provided by these authors refers to removing positively charged calcium ions and loosening membrane structure for it by better penetration by extra calcium and stimulates cell metabolism. According to [57] the assistance of the magnetic field boosts the fermentation process and avoid addition of commonly used additives (normally: $1\%$ toluene, $4\%$ ethanol and $0.075\%$ triton X-100) to enhance the permeability of the of the cell membranes.

The recent study of Santos et al. [58] considered static magnetic field assisted biomass and glutathione

TABLE IV.
MAIN EXPERIMENTS PERFORMED ON MAGNETIC FIELD ASSISTED *S. cerevisiae* FERMENTATION

| Field | Reactor and Culture medium | duration | Ref. |
|---|---|---|---|
| Static Magnetic field Axial, $150 mT$ created by a solenoid entirely encircling a 7.5 litre BioFlo III fermentor (New Brunswick) | 7.5 litre BioFlo III fermentor, New Brunswick Initial sugar concentration: $200 g/L$ | $16h$ | Galonja-Coghill et al. [57] |
| A solenoid $8.2 cm ID$ and $15 cm$ height $100 kHz$ $B = 250 mG$ | $400 mL$ reactor Molasses (2005a)(2005b) | $25h$ | Zapata et al. [59], [60] |
| Static magnetic field $25.0 - 34.3 mT$ Helmholtz pair ($10 cm ID$ and $18 cm$ external diameter) | $250 mL$ reactor | $72h$ | Santos et al. [58] |

(GSH) production. In general, there are no novelties in this study except the product (GSH), of course. The data are hard to be analyzed since no direct experimental information is provided. The main approach is to present outcomes of planned experiments treated by the ANOVA analysis. This simply means a low number of experiments (only 7 runs) behind the studies and problems to extract tendencies in the biological system behaviour when a magnetic field is imposed. The quantitative results obtained for runs lasting $24, 48$ and $72h$, and field exposures from 8 to $16h$ range (of each run), reveal that:

a) The biomass results (first point taken after $24h$ of incubation) show field effects induced augmentations with respect to the control run.
b) The higher biomass concentration ($16.26 g/L$) was obtained with a $72h$ fermentation run and the longest field exposure time of $16h$ at $25.0 mT$. Similarly, in a run with exposure time of 8h, and $25.0 mT$, the biomass was almost the same, $16.11 g/L$. The authors attributed the differences to the time of the incubation period ($72h$ and $48h$, respectively), but based on the previous comments in this section of the review we have to mention that the time for exposure coincides with the exponential phase. Therefore, the results are like those in the other studies commented here, for example those of Motta [48], [49], [54], [55].
c) With respect to the GSH yield, a stimulatory effect was observed in all the runs except those exposed for $8h$ and $12h$ in fields of $34.3 mT$ and $29.6 mT$, respectively. The specific GSH yield varied from 15.68 to 20.91 $mg$ GSH/gbiomass at $72h$ of incubation: the highest value was obtained with the longest exposure time ($16h$) and the lowest magnetic field induction ($25.0\,mT$).

*V.2. Low-intensity and high frequency fields*

The experiments of Zapata et al.[59] with $100\,kHz$ in molasses as a substrate reveal that only $200\,s$ exposures affect the growth kinetic by increase of about $20\%$ the growth rate, and up to $28\%$ of the yield, as well. At the same time, the saturation constant in the Monod model reduces with $26\%$. In general, different operating modes were applied: A-mode, with magnetic treatment prior to the fermentation and without aeration; B-mode, with magnetic treatment prior to the fermentation and aeration; C-mode, with aeration and magnetic treatment prior to the fermentation as first run and after the lag phase (after $4h$); D-mode, similar to C-mode but field application during the lag phase (at $2h$ and $4h$ after the fermentation onset). The parameters of the Monod model obtained by Zapata et al. (2005) are summarized in Table **V** According to those data the best process performance was obtained with B-mode that could be attributed to an initial heating of the substrate (molasses) and initiation (boosting) of the process of fermentation.

TABLE V
DATA OF ZAPATA ET AL. [59] ON FERMENTATION IN MOLASSES ASSISTED BY $100\,kHz$ $B=250\,mG$. MONOD' MODEL PARAMETERS IN DIFFERENT OPERATING MODES

| Monod' model parameters | | | | | | |
|---|---|---|---|---|---|---|
| Mode | $\mu_{max}\,(h^{-1})$ | | $K_s\,(g/L)$ | | Yield coefficient $Y_{xs}\,(g/g)$ | |
|  | E | C | E | C | E | C |
| A | 0.41 | 0.36 | 34 | 27 | 0.11 | 0.11 |
| B | 0.24 | 0.24 | 61 | 83 | 0.18 | 0.14 |
| C | 0.29 | 0.26 | 58 | 50 | 0.15 | 0.15 |
| D | 0.36 | 0.30 | 74 | 64 | 0.26 | 0.23 |
| **E**- exposed ; **C**-control | | | | | | |

*V.3. High-intensity static magnetic field*

Growth and fermentation activity experiments of Anton-Lefebvre et al.[[29] with yeasts pre-exposed cultures to $4x20T$ pulses in rapid sequence and monitored up to $6h$ (samples at every $30\,min$) in a rotary shaker did not provide results indicating magnetic field effects on both the growth rate and the ethanol production rate, and the glycerol production, as well. The authors have tried to explain the results with some models [51], [61] concerning yeast behaviour under fields up to $14T$, commented briefly next.

The experiments of Iwasaka [51] and Ikehata [61] revealed deceleration in the cell growth and this effect was reproducible in all runs of proliferation. Furthermore, the gas ($CO_2$) produced by the cell decreased at high magnetic fields of order of $10T$. Further, in a gradient magnetic field with a vertical orientation (parallel to the gravity), the field effect on the cell proliferation was also towards deceleration. Moreover, they suggested some mechanisms explaining the inhibitory effect of the field commented further in this article.

*V.4. Fermentation with a recirculation loop and external magnetization*

In contrast to the common approach concerning the working volume for proliferation or fermentation experiments to entire encircled by the magnetic system, as in the dominating studies, discussed in this review, there exists an approach [62] employing intermittent, periodic magnetization by a closed loop including the reactors and a small chamber for magnetic exposure. The culture medium passes through the magnetization chamber and enters the reactor again. The residence time for magnetization depends on intensity of circulation, the cross section of the exposure chamber, and its length. Detailed analysis of this type of fermentors was performed recently elsewhere [2].

Ethanol fermentation by *Saccharomyces cerevisiae* in such a reactor parameters was performed an the outcomes analyzed by ANOVA. The analysis of experimental data led to the conclusion that cell suspension may involve an electric field or magnetic and electric fields induced in the conducting culture and alter the metabolic activity through the micro-level dynamo (MLD) concept [2] commented further in this review.

The ethanol production varied with the recycle velocity showing two positive responses for $0.9$ and $1.2\,m/s$. The yield of ethanol at $0.6\,m/s$ and $1.2\,m/s$ was almost the same ($\approx 4.2\,(g.L^{-1}.h^{-1})$), but lower that that observed at $0.9\,m/s$ and $1.2\,m/s$ ($\approx 5.75\,(g.L^{-1}.h^{-1})$). The difference in $12\%$ was considered as "a window effect" with respect to the contribution of the circulation velocity to the total process performance. The latter, was attributed also to the field effect, i.e. frequency of magnetization.

The crucial element in this reactor design is the "circulation loop". As it was commented [2] that a reduction in the loop internal diameter increases the circulation velocity and reduces the time of magnetization. However, this causes enormous mechanical stresses in the culture medium that reduce the efficiency of the fermentation with respect to reactor with a stirring blade only. Due to strong mechanical stresses the cells become almost dormant, both the growth and the metabolic activity reduce below the level of the conventional bioreactor. Hence, the recirculation loop has to be designed with caution and balancing between the benefits of the external magnetization (easily resolved scale-up problems, for example) and the effect of the additional mechanical stresses in the loop.

# VI. Discussion and Further Analysis

*V.1. Common explanations still exiting in the literature*

Despite the many positive results no definitive explanations were developed on the basis of the experiments performed so far. Several concepts were referred, among them: changes in the transmembrane voltage, $Ca^{2+}$ distribution, enzyme turnover rate owing to local oscillation of the ion concentration (the ion activation model [11]) or to influence of their radical reactions [63] in addition to dipole polarization of membrane components enhanced $Ca^{2+}$ in flux, etc. No final discrimination of the effects and mechanisms was done.

*VI.1.1. Growth and fermentation activity: some physical models [51], [61]-high static fields*

In the context of high static magnetic field effects on the proliferation of *Saccharomyces cerevisiae* we have to mention two detailed studies [51], [61], both of them performed under high fields up to $14T$. These authors observed deceleration in the cell growth and this effect was reproducible in all runs of proliferation. Furthermore, the gas ($CO_2$) produced by the cell decreased at high magnetic fields of order of $10T$. Further, in a gradient magnetic field with a vertical orientation (parallel to the gravity), the field effect on the cell proliferation was also towards deceleration. Moreover, they suggested some mechanisms explaining the inhibitory effect of the field, namely [51], briefly outlined next (see [2] for detailed analysis):

- **Mechanism-1** considers a concentration gradient of the oxygen because its concentration beneath the air-solution boundary is higher than in the bottom part of the vessel. However, the results indicated that the gradient magnetic fields decelerated the proliferation of the yeast.
- **Mechanism-2** focuses on magnetic forces acting on diamagnetic water molecules and acrostically appearing as hydrostatic forces. The diamagnetic force acting on the condensed $CO_2$ gas adds to the existing one and acts on the yeast sedimentation.
- **Mechanism 3** suggests that the paramagnetic oxygen molecules are driven by the magnetic gradient thus provoking a macroscopic fluid convection.

As it was commented recently [2] these mechanisms seems reasonable but they do not support the observations. In conclusion, yeasts under gradient magnetic fields ($14T$) exhibited decelerated growths and Iwasaka et al. [51] attributed the negative effects to the transport of the paramagnetic oxygen gas and the diamagnetic materials (carbon dioxide, water, etc.) and mainly due to the decreased oxygen supply. However, these mechanisms [51], [61] will never work if a real fermentation with a mechanical mixing is carried out.

*VI.2. Micro-level dynamo concept*

*VI.2.1. MLD to proliferation experiments*

The results presented on *Saccharomyces cerevisiae* proliferation in magnetic fields reversal claim that either the exposures inhibit the development [27] or there is no effect on the cells [29]. The results of the latter studies confirm those of:
- Ruiz-Gomez et al [31] with field of $0.35 - 2.5 mT$ and $50 Hz$.
- Nakasono et al. [64], with field of $< 300 mT$ and $50 Hz$ for $24h$.
- Ikehata et al. [61], with a static field up to $14T$, where $5900$ genes were examined.
- Iwasaka et al. [51], with a static field up to $14T$ and observations on cell proliferation
- Binninger and Ungvichian [65], with $60 Hz$, $20 \mu T$ with no effect either on the genes expression or cell division.

The missing magnetic field effect was attributed [29] to the fact that in the dominating cases it was studied in the *stationary phase* of growth where the cells have already attained resistance mechanisms to stress. The failure to find magnetic field effects in most of the experiments is mainly due to this preliminary defined condition. The experiments of Anton-Leberre et al. [29] were conducted in the *exponential phase* with not yet developed resistance mechanism of the cells, no limitation to the nutrient supply and an active phase of growth.

An important parameter is the energy absorbed during the magnetic field exposure. In static fields, in general, there is no field energy transmission to the living cells, but the field can alter the energy levels of certain molecules [29], 66]. However, pulsed and time-varying fields (mainly $50 Hz$ in this analysis) induce eddy currents in the aqueous culture media, which are highly conductive due to additional ions added. In other words, this is a Faraday coupling but with a special emphasis of the magneto-hydrodynamic phenomena in the conducting liquids in both sides the cell membrane (outside and inside the cells). These eddy currents generally produce heating in the culture medium and the cell cytoplasm. However, we have to recall that the penetration of the field-induced eddy currents is limited to the surface of the liquid only, the so-called "skin effect", that means: only those cells which are close to air-culture interface are affected. The eddy currents affected zone decays rapidly in depth with increase in both the field magnitude and the frequency. With high intensity magnetic pulses [29], the field cannot penetrate the sample in depth in contrast to the case where only static field is applied.

Besides, the eddy currents create their own magnetic field that in a co-action with the principle one (that used for the exposure) creates small dynamos mixing the liquid at micro level. Such microscopic dynamos may enhance the mass transfer at the vicinity of the cell surfaces where stagnant liquid layers exist and the mass transfer is totally controlled by diffusion, irrespective of the intensity of mechanical mixing. Since, in most of the proliferation experiments, the mixing was not intensive and the cell growth was already accomplished, it might be expected that no field effects boosting the cell living activity were found. Any suppression effects [27] might be attributed to the local heating that alters the conditions of normal cell growth, but in general, it cannot stop the process due to rapid heat dissipation in the medium and in the vessel.

In contrast to the above mentioned results [27], [29], the experiments of Motta et al. [48], [49], [54], [55] show just the opposite tendencies, i.e. stimulated biomass growth in exposed cultures. The answers to the questions emerging are:

- The exposure of a static conduction medium to time-varying magnetic fields (pulsed or AC) or agitation of the same suspension in a static magnetic field, from the point of view of eddy current generation, is one and the same, because there is a relative motion and variation in time of the medium and the magnetic flux lines: that is there is a Faraday coupling between the liquid phase and the magnetic field.
- The MLDs works in the vicinity of the cell-liquid interface thus enhancing the mass transfer by reducing the diffusion mass transfer resistance, i.e. there is an increase of the nutrient flux to the cells, which without any changes in the morphology and the metabolism augments their growth rate. From this point of view, the experiments of Medhindu and Berg [27] in the "positive windows " (with positive field effects) in accordance with their terminology are close to those of Motta [48], [49] . That is, in the low field intensities and frequencies used in those experiments allow the eddy currents to penetrate the liquid in depth and the MLDs will be at work.
- When strong static fields are applied [29] no eddy currents emerge, except at the very beginning when the field is turned on. Then, the field has no effect on the transport processes in the stagnant liquid. If the field is strong and pulsed or static, and the liquid is agitated, the eddy currents do not penetrate into the bulk, this is the so-called "skin effect".

Hence, the MLDs cannot affect the transport process in bulk of the cell suspension.

*VI.2.2. MLD to fermentation studies*

From the position of the MLD concept the positive outcomes of the magnetically exposed fermentation could be explained, namely:

- The inhibitory effect of the ethanol on the *S. cerevisiae* reproduction is the same in both the exposed and the unexposed process because the macroscopic mixings created by the agitators lead to almost equal conditions allowing the ethanol to be removed from the vicinity of the cell surfaces.
- With macro mixing, this can be attained by increasing the rotation of the stirrer to some extent, but at the vicinity of the cell surface there is a thin liquid layer which is almost stagnant and close to saturation with respect to the ethanol released by the cell. Hence, this layer becomes the main mass transfer resistance, namely: with respect to the nutrients from the liquid to the cells and from the cells to the liquid, with respect to the metabolic products, as well.
- Hence, the increased nutrient flux towards the cells is augmented by the micro dynamos that with unchanged cell metabolisms enhance their growth and production.
- Moreover, the micro mixing due to micro dynamos at the vicinity of the cell/liquid interface augments the removal of the ethanol, unlike the case of unexposed process where this removal is controlled by pure molecular diffusion only. Hence, the ethanol inhibitory effect at the vicinity of the cell-liquid interface can be reduced to some extent.
- Besides, we may only suggest, without any data at the moment, that eddy currents at the vicinity of the cell walls may stimulate the transport across the membranes as it was mentioned above in accordance with dozen of hypotheses existing in the literature. However, this suggestion needs further experimental verifications.

*VI.2.3. Main points in the MLD concept  
and some notes*

The hypothesis of micro-level dynamos [2] refers to the liquid-side mass transfer in the cell growth process and to only mechanical effects in the liquid phase. The MLD concept does not touch the intracellular process and tries to arrange the existing experimental facts into a common structure. It gives explanations of the magnetic field effects within the framework of the macroscopic (extracellular) mass transfer processes, without any relations to "magic magnetic effects". The effects on the cell membranes are encompassed by MLD concept, too, but the existence of electric currents near the cell (eddy currents) with their own magnetic fields and potentials suggests effects on all phenomena mentioned above, as $C^{2+}$ influx, ion activity, enzyme turnover, etc. We have to recall that all these effects take place at the near-field range, at the vicinity of the cell membrane interface. Hence, the magnetic field effects have to be looked for at this location irrespectively of the scale of the magnetic field application: over Petri dishes or in large magnetically assisted bioreactors. In this context, recall the models of Iwasaka et al. [51] concerning also magnetically controlled transport processes outside the cell. However, the proposed MLD's physical mechanism is quite different and works at a micro-level near the cells in contrast to the models of Iwasaka explaining the macro-mixing in the system.

In the context of some biological effects as a result to the exposure to magnetic fields, we have to refer to [67] noting that the Larmor precession relevant to the intracellular diffusion processes requires a minimum field intensity to provoke some biological effects of about $10^{13}$ *Gauss* (fields impossible in conventional biotechnological laboratories).

*VI.2.4. MLD to heat dissipation and heat shock response of the yeasts: a hypothesis*

This last section of the article address a hypothesis relating physical and biological phenomena affecting the cell life. Especially, we try to relate the heat-shock response of the cells and the thermal phenomena in the both the intracellular and the extracellular liquid and the relevant heat dissipation mechanisms.

**Yeast responds to heat shocks** by a complicated response in order to adapt and survive. This response includes rapid alterations in global transcription and metabolic remodeling [68], [69]. In this context, the heat shock protein (HSP) genes have been known to have a vital role in coping with heat shock. Many HSPs function as molecular chaperones for protecting unfolded or unstable proteins from degradation or aggregation [70]. The heat shock response also involves the synthesis of metabolic enzymes and antioxidant defense proteins. The production of many of these proteins can be provoked by a variety of stresses [71]. Most HSPs act as chaperones, either directing folding and assembly of proteins or dissociating aggregates [72]. Nevertheless HSPs are among the most highly conserved proteins in evolution, with respect to Hsp104 among them [72] for example, the lethality associated with heat shock is enhanced up to 1000 fold in *S. cerevisiae* lacking this protein [73],[74].

In *Saccharomyces cerevisiae*, this response is controlled by the heat shock transcription factor *Hsf1p*, which binds the heat shock response element (HSE) [75], [75]. As a consequence, the level of the disaccharide trehalose increases rapidly and dramatically after heat shock, along with the induction of *HSPs* [76]. Trehalose has been shown to stabilize the structures and enzymatic activities of proteins against thermal denaturation in vitro. Trehalose can prevent the aggregation of misfolded proteins [78], and serves as a positive regulator of the *Hsf1p* activity, suggesting that trehalose functions to protect proteins at the initial stages of the heat shock response before *HSPs* have been fully induced [79]. In addition, many oxidative stress-responsive genes are involved in the heat shock response [71], [[80], suggesting that these genes may have important roles in coping with different types of stress. The results of Ye et al. [69] showed that the heat-stressed *S. cerevisiae* cell accumulated trehalose and glycogen, which protect cellular proteins against denaturation, and modulate its phospholipid structure to sustain stability of the cell wall. Further, the increased temperature enhances the consumption of oxygen in respiration, and heat shock may enhance the generation of reactive oxygen species (ROS) in mitochondria caused by an enhanced respiration rate [81].

According to Ye et al. [69], the response logic of *S. cerevisiae* cells to heat shock involves both the gene expression profiles and the metabolic pathways. That is, the yeast cells confront with heat stress and accumulate trehalose and glycogen [79], [82], [83]. A strong correlation between trehalose content and stress resistance has been demonstrated for a variety of stresses [78], [79], [82] such as heat, osmotic stress, and ethanol. This leads to protection of the cellular proteins and alternation of the phospholipid structure that sustains the cell wall stability. Consequently, the synthesis of amino acids and nucleotides is repressed due to increased energy requirement as result of the changed protein metabolism and modification. In this chain, the enhanced respiration leads to oxidative stress.

Hence, the evolution of metabolic systems allowing adaptation to sudden stresses could occur by production of stress protectants in addition to providing glycolytic safety valves. In this context both glycerol and trehalose appear to be ideal candidates [83] (Li et al., 2009). One of the consequences of sudden stress is a dramatic decrease in the ATP demand from macromolecular biosynthesis [83], [84] suggested, on the basis of the glycolysis model, that under conditions of growth arrest, with low ATP demand, cells face the threat of substrate accelerated death.

Taking into account the complex role of glycerol in the yeast cell life, Ferreira and Lucas [85] suggested that the *S. cerevisiae* Stl1 defective strain, for instance, is a glycerol-related mutant, displaying a growth limitation at high temperature, both on fermentable or respiratory conditions. Moreover, these authors suggested that Stl1p contributes to the fine-tuning of glycerol internal levels, applying to heat stresses and the cells go through a switch from fermentative to respiratory metabolism.

Additionally, the yeast *S. cerevisiae cell* wall comprising a 10 nm thick layer of polysaccharides and proteins is not static and dynamically reacts to changes in the environment. According to Karreman et al. [86], the presence of the stress response protein Hsp12p resulted in a more flexible cell, thereby confirming our hypothesis that Hsp12p acts a cell wall plasticizer in the yeast *S. cerevisiae*.

**The heat dissipation** in liquid bodies involves two main mechanisms: **conduction** and **convection**. After the preamble concerning the heat shock response of cells, let us try to suggest the relationship through the MLD and the heat-shock response of the yeast. The MLDs, which is a Faraday coupling to the conducting liquids, both intracellular and extracellular (the culture medium) and acting at micro-level, have two principle physical effects, namely:
- Mixing as a result of the toques on the liquid volumes created by the principle (external) magnetic field and that of the eddy currents.
- Heating, in fact, ohmic heating of the liquids volumes and consequent heat dissipation

However, the scale defines what is the main transport mechanism concerning the mixing and the heat dissipations. Because the cells are extremely small in size, then all transport in the intracellular volume are due to diffusion: **diffusional mass transfer**, **creeping fluid flow** and **heat transport by conduction**. This limitation due the length scale (cell size) does not exist outside the cells (extracellular liquid). The latter means, that either the macroscopic mixing due the stirrer or that in the vicinity of the cell membranes (the external liquid side) makes possible transport by convection: both mass transport and heat dissipation.

Now, let us look at the main conditions in the proliferation experiments and the fermentations discussed in this review. The proliferation experiments are generally carried out in Petri dishes or small ampoules that avoid any macro-mixing in the culture medium. Therefore, all transports in the culture medium and inside the cells are based on diffusion. When the eddy currents heat the intracellular liquid the only possible mechanism is the heat conduction towards and membrane. This mechanism cannot be enhanced by either MLD due to the scale limitation not by other physical tools. Consequently, the heat dissipation through the membrane and then to the culture medium is also by heat conduction. Therefore, in the proliferation experiments there are limited possibilities to dissipate the heat generated inside the cells. This might be a reason for a heat shock and all relevant phenomena mentioned above, but performing themselves at a macroscopic level as reduced cell growth. The MLD induced mixing in the vicinity of the cell membrane is not enough to enhance the heat removal because the entire culture medium is almost stagnant.

When fermentation is performed, we have well developed macro-mixing equalizing the temperature over the entire reactor volume. This mixing, combined with the MLD induced one is more effective in heat removal from the cells towards the culture medium. Both effects, leads to cell growth and enhanced metabolic activity.

This hypothesis relating the MLD concept and the heat removal from the cells try to explain that in the proliferation experiments the cells are more exerted to heat shocks than in the fermentation runs. As consequences, we have different behaviours when magnetic fields are imposed. Obviously, there thermal and no-thermal effects on the cells irradiated by magnetic field. The principle question is: Which of them dominates and control the cell life? The answer is not straightforward and needs a detailed analysis of the phenomena and adequate modelling. The latter means scale analysis, evaluation of scale effects, order of magnitudes, etc. This figures future studies and analyses requiring a lot of experimental a data that to some extent are still missing in the literature. However, it might suggest that the above outlined hypothesis would provoke studies in that direction.

## VII. Final Comments

The present analysis of magnetic field effects on *Saccharomyces Cerevisiae* can be briefly outlined in two major groups of effects [2], of magnetically assisted processes, namely:

**A) Advantages**
- There are positive windows in the cell proliferation at low-frequency and low-intensity field exposures.
- High static fields suppress the cell growth and proliferation.
- The fermentations in batch reactors with mixing (experiments of Motta et al. [48],[49],[54,[55]) and those of Perez et al. [62] strongly indicate that there are also positive windows when static and pulsed fields are applied, respectively.

**B) Disadvantages**
- Most proliferation studies do not give answers or suggest ideas how the results can be developed towards large scales.
- When fermentations are performed, most of the studies concerning proliferations as initial steps do not extend the explanation to the problems at macro level.
- The living microorganisms in bioreactors are exposed to many rough conditions because, in fact, the conditions do not replicate exactly their natural environments. Mixing, overgasing, changing pH, strong shear stresses, are common in laboratory and industrial fermentors worldwide.

- Magnetic field exposures contribute to stresses imposed on the cells and bacteria changing their natural habit of life. Hence, magnetically assisted reactors create too many stresses to the living microorganism and the route to the best reactor performance passes through careful preliminary analyses, correct design and systematically carried out experiments.

The mentioned outcomes just figure what has to be done and what has to be avoided. The development of these ideas needs interdisciplinary knowledge at the edge of fluid mechanics, cell biology, reactor design and magnetism (at micro level and an engineering design, too). Prior to start your own research on a magnetically assisted free-cells bioreactor and reading articles on macroscopic fermentation experiments, some important sources of information are recommended, among them: Blank and Goodman [87], Pazur et al. [88], Hunt et al. [1] and Hristov [2].

**Acknowledgements**

The support of FAPERJ (Rio de Janeiro, Brazil) to the author via a grant ( June –September 2010) auxiliary to the project "*Development of Technologies with Electromagnetic Fields Applied to Process Engineering*" seeking to the biodiesel production by unconventional methods using immobilized cells in bioreactors assisted by electromagnetic field" , (Ref. Proc. FAPERJ Nº 110.816/2008)" is highly appreciated. The author (JH) also thanks Prof, Viara Ivanova (University of Food Technologies, Plovdiv, Bulgaria) for valuable comments and Prof. J.M. Valverde (University of Sevilla, Spain) for help with copies many old articles.

## References

[1]   R. W. Hunt, A.Zavalin, A. Ashish Bhatnagar, S. Chinnasamy, K. C. Das, Electromagnetic Biostimulation of Living Cultures for Biotechnology, Biofuel and Bioenergy Applications , *International Journal of Molecular Sciences*, 10 *(2009)* 4515-4558. doi:10.3390/ijms10104515

[2]   J. Hristov, Magnetic field assisted fluidization - A unified approach. Part 8. Mass Transfer: magnetically assisted bioprocesses, *Reviews in Chemical Engineering, vol.* 26 *(3-4) (2010)*, 55-128, doi: 10.1515/revce.2010.006 .

[3]   J. Hristov, V. Ivanova, Magnetic Field Assisted Bioreactors, in: *Resent Research Developments in Fermentation and Bioengineering* 2 *(1999)* 41-95.

[4]   G. S. Silva, P. C. M. da Rós, J. C. Santos, V .H. Perez, H .F. de Castro, Enzymatic Transformation of Palm Oil in Biodiesel using Porcine Pancreatic Lipase Immobilized on Hybrid Matrix, *International Review of Chemical Engineering (IRECHE)*, 1 *(2009)* 609-613.

[5]   R. H. W. Funk, T. Monsees, N. Özkucur, Electromagnetic effects –From Cell biology to medicine, *Progress in Histochemistry and Cytochemistry*, 43 *(2009)* 177-264. doi:10.1016/j.proghi.2008.07.001.

[6]   A. Markkanen, J. Juutilainen, J. S. Lang, J. Pelkonen, T. Rytimaa, J. Naarala, Effects of 50 Hz Magnetic Field on Cell Cycle Kinetics and the Colony Forming Ability of BuddingYeast Exposed to Ultraviolet Radiation, *Bioelectromagnetics* 22 *(2001)* 345-350.

[7]   A. M. Carr, Radiation checkpoints in model systems, Int *J Rad. Biol.* 66 *(1994)* 133–139.

[8]   A. M. Woodward, D. B. Kell, On the nonlinear dielectric properties of biological systems- *Saccharomyces cerevisiae, Bioelectrochemistry and Bioenergetics* 24 *(1990)* 83-100.

[9]   Tatabe W. A. M. Woodward and D. B. Kell, Nonlinear dielectric properties and their relationship to the physiological state (resting or active) of *Saccharomyces cerevisiae*, Mem. Fac Eng. Osaka City University, 35 *(1994)* 53-57.

[10]  L. M. Mir, M. F. Bureau, J. Gehl, R. Rangara, D. Rouy, J.M.Caillaud, P. Delaere, D. Branellec, B. Schwartz, D. Scherman, High-efficiency gene transfer into skeletal muscle mediated by electric pulses. *Proc NatlAcad Sci USA*, 96 *(1999)* 4262–4267.

[11]  J. Schimmelpfeng, H. Dertinger, Action of a 50 Hz magnetic field on proliferation of cells in culture, *Bioelectromagnetics 18 (1997)* 177-183.

[12]  M. R. Cho, H. S. Thatte, R. C. Lee, D. E. Golan, Reorganization of microfilament structure induced by AC electric fields. *FASEB J 10 (1996)* 1552–1558.

[13]  A. A. Pilla, Weak time-varying and static magnetic fields: from mechanisms to therapeutic applications. In: Stavroulakis P, editor. Biological effects of electromagnetic fields: mechanisms, modeling, biological effects, therapeutic effects, international standards, exposure criteria. Berlin: Springer (2003)

[14]  M. S. Markov, Magnetic field therapy: a review, *Electromagn. Biol. Med.*, *26 (2007)* 1–23.

[15]  D. J. Muehsam, A. A. Pilla, Lorentz approach to static magnetic field effects on bound-ion dynamics and binding kinetics: thermal noise considerations, *Bioelectromagnetics*, *17 (1996)* 89–99.

[16]  L. C. Lipus, J. Krope, L. Crepinsek, Dispersion Destabilization in Magnetic Water Treatment, *J. Colloid and Interface Science* 236 (2001) 60–66.

[17]  C. Y. Tai , C-K. Wu, M-C. Chang, Effects of magnetic field on the crystallization of $CaCO_3$ using permanent magnets, *Chem. Eng. Sci.* 63 (2008) 5606-5612. doi:10.1016/j.ces.2008.08.004.

[18]  C. Y. Tai , M-C. Chang, R-J. Shieh, T.G. Chen, (2008b) , Magnetic effects on crystal growth rate of calcite in a constant-composition environment, *J.Crystal Growth* 310 *(2008)* 3690–3697. doi:10.1016/j.jcrysgro.2008.05.024 .

[19]  L. Lahuerta Zamora , G. M. Antón Fos, P. A. Alemán López, R. V. Martín Algarra, Magnetized Water: Science or Fraud?, *J. Chemical Education* 85 (2008) 1416- 1418.

[20]  A. D. Rosen, Mechanism of action of moderate-intensity static magnetic field on biological systems, Cell Biochem Biophys. 39 *(2003)* 163-173.

[21]  S. N. Batchelor, S. W. Kay, K. A. McLauchlan, I. A. Shkrob, Time-resolved and modulation methods in the study of the effects of magnetic fields on the yields of free-radical reactions, *J. Phys. Chem.,* 97 *(1993)* 13250-13258.

[22]  K. A. McLauchlan, Are the environmental magnetic fields dangerous? , *Physiscs World, 92 (1992)* 41-45.

[23]  M.Cifra, J.Z.Fields, A. Farhadi, Electromagnetic cellular interactions, *Progress in Biophysics and Molecular Biology*, *(2010)* Article in Press . doi:10.1016/j.pbiomolbio.2010.07.003

[24]  T. Y. Tsong, Molecular recognition and processing of periodic signals in cells: study of activation of membrane ATPases by alternating electric fields, *Biochimica et Biophysica Acta (BBA) - Reviews on Biomembranes* 1113 *(1992)* 53-70. doi:10.1016/0304-4157(92)90034-8 .

[25]  K. R. Foster, H. P. Schwan, Dielectric properties of tissues and biological materials: a critical review, *Critical Reviews in Biomedical Engineering* 17 *(1989)* 25-104.

[26]  N. Stoykov, J. Jerome, L. Pierce, A. Taflove, Computational modeling evidence of a nonthermal electromagnetic interaction mechanism with living cells: microwave nonlinearity in the cellular sodium ion channel, *IEEE Transactions on Microwave Theory and Techniques* 52 *(2004)* 2040-2045.

[27] M. Mehedintu, H. Berg, Proliferation response of yeast *Saccharomyces cerevisiae* on electromagnetic field parameters, *Bioelectrochemistry and Bioenergetics 43 (1997)* 67-70.

[28] U. Fiedler, U. Grobner, H. Berg, Electrostimulation of yeast proliferation. *Bioelectrochemistry and Bioenergetics 38 (1995)* 423–425.

[29] V. Anton-Leberre, E. Haanappel, N. Marsaud, L. Trouilh, L. Benbadis, H. Boucherie, S. Massou, J. M. Francois, Exposure to High Static or Pulsed Magnetic Fields Does Not Affect Cellular Processes in the Yeast *Saccharomyces cerevisiae*, *Bioelectromagnetics* 31 *(2010)* 28-38. doi 10.1002/bem.20523.

[30] J. Novak, L. Strasak, L. Fojt, I. Slaninová, V. Vetterl, Effects of low-frequency magnetic fields on the viability of yeast *Saccharomyces cerevisiae*, *Bioelectrochemistry 70 (2007)* 115–121.

[31] M. J. Ruiz-Gomez, M. I. Prieto-Barcia, E. Ristori-Bogajo, M. Martinez-Morillo, Static and 50 Hz magnetic fields of 0.35 and 2.45mT have no effect on the growth of *Saccharomyces cerevisiae*, *Bioelectrochemistry 64 (2004)* 151–155.

[32] M. J. Ruiz-Gomez, F. Sendra-Portero, M. Martinez-Morillo, Effect of 2.45 mT sinusoidal 50 Hz magnetic field on *Saccharomyces cerevisiae* strains deficient in DNA strand breaks repair, *Int. J. Radiat. Biol. 86 (2010)* 602–611.

[33] E. Gülbandilar, Effect of pulsing electromagnetic field on the growth of *Saccharomyces cerevisiae*, *Annals of the Dumlupinar University (Turkey)*, 9 *(2005)* 55-64.

[34] L. Fojt, L. Strasak, V. Vetterl, J. Smarda, Comparison of the low-frequency magnetic field effects on bacteria *Escherichia coli*, *Leclercia adecarboxylata* and *Staphylococcus aureus*, *Bioelectrochemistry 63 (2004)* 337 – 341.

[35] L. Strasak, V. Vetterl, J. Smarda, Effects of low-frequency magnetic fields on the bacteria *Escherichia coli*, *Bioelectrochemistry and . Bioenergetics 55 (2002)* 161-164.

[36] C. Luceri, C. De Filippo, L. Giovannelli, M. Blangiardo, D. Cavalieri, F. Alietti, M. Pampaloni, D. Andreuccetti, L. Pieri, F. Bambi, A. Biggeri, P. Dolara, Extremely lowfrequency electromagnetic fields do not affect DNA damage and gene expression profiles of yeast and human lymphocytes, *Radiation Research 164 (2005)* 277–285.

[37] K. Simizu, Y. Nakaoka, T. Yamamoto, High density 60Hz magnetic field has no effect on mutagenicity and growth of budding years *Saccharomyces cerevisiae*, *Radiation Safety Management*, 1 *(2002)* 17-20.

[38] P. F. Verhulst, Notice sur la loi que la population suit dans son accroissement, *Correspondence mathematique et physique, A. Quetelet* (Brussels) *10 (1838)* 113-121.

[39] E. S. Otabe, S. Kuroki, J. Nikawa, Y.Matsumoto, T. Ooba, K. Kiso, H. Hayashi, Yeast cells proliferation on various strong static magnetic felds and temperatures, *Journal of Physics: Conference Series 156 (2009)* 012016. doi:10.1088/1742-6596/156/1/012016 .

[40] E. Neumann, Digression on chemical electromagnetic field effects in membrane signal transduction cooperatively paradigm of the acetylcholine receptor, *Bioelectrochemistry, 52 (2000)* 43-49. doi:10.1016/S0302-4598(00)00082-9 .

[41] D. Crouzier, A. Perrin, G. Torres, V. Dabouis, J.-C. Debouzy, Pulsed electromagnetic field at 9.71 GHz increase free radical production in yeast (Saccharomyces cerevisiae), *Pathologie Biologie 57 (2009)* 245–251. doi:10.1016/j.patbio.2007.12.003

[42] P. Gos, B. Eicher, J. Kohli, W. D. Heyer, Extremely High Frequency Electromagnetic Fields at Low Power Density Do Not Affect the Division of Exponential Phase *Saccharomyces cerevisiae* Cells, *Bioelectromagnetics 18 (1997)* 142–155.

[43] W. Grundler, Intensity- and frequency-dependent effects of microwaves on cell growth rates, *Bioelectrochemisty and Bioenergetics*, 27 *(1992)* 361–365.

[44] I. Vrhovac, R. Hrascan, J. Franekic, Effect of 905 MHz microwave radiation on colony growth of the yeast *Saccharomyces cerevisiae* strains FF18733, FF1481 and D7, *Radiology and Oncology 44 (2010)* 131-134. doi:10.2478/v10019-010-0019-7.

[45] P. Gos, B. Eicher, J. Kohli, W. D. Heyer, Nomutagenic or recombinogenic effects of mobile phone fields at 900 MHz detected in the yeast *Saccharomyces cerevisiae*, *Bioelectromagnetics, 21 (2000)* 515-23.

[46] A. Usatii, E. Molodoi, A. Rotaru, T. Moldoveanu, The influence of low intensity millimeter waves on the multiplication and biosynthetic activity of *Saccharomyces carlsbergensis cnmn-y-15* yeast, *Annals of the University of Oradea -Faculty of Biology (Romania) 17 (2010)* 208-212.

[47] F. E. Van Nostran, R. J. Reynolds, H. G. Hedrick, Effects of high magnetic field at different osmotic pressures and temperatures on multiplication of *Saccharomyces cerevisiae*, *Applied Microbiology 15 (1967)* 561-563.

[48] J. B. Muniz, M. Marcelino, M. da Motta, A. Schuler, M. A. da Motta, Influence of static magnetic field on *S. cerevisae* biomass growth, *Brazilian Archives of Biology and Technology 50 (2007)* 515-520.

[49] M. A. Motta, E. J. N. Montenegro, T. L. M. Stamford, A. R. Silva, F. R. Silva, Changes in *Saccharomyces cerevisiae* development induced by magnetic fields, *Biotechnology Progress 17 (2001)* 970-973.

[50] J. Blanchard, C. Blackman, Clarification and application of an ion parametric resonance model for magnetic field interactions with biological systems, *Bioelectromagnetics 5 (1994)* 217-238.

[51] M. Iwasaka, M.Ikehata, J. Miyakoshi, S.Ueno, Strong static magnetic field effects on yeast proliferation and distribution, *Bioelectrochemistry 65 (2004)* 59-68.

[52] C. B. Coleman, R. A. Gonzalez-Villalobos, P. L. Allen, K. Johanson, K. Guevorkian, J. M. Valles, T. G. Hammond, Diamagnetic levitation changes growth, cell cycle and gene expression of *Saccharomyces cerevisiae*, *Biotechnology and Bioengineering 98 (2007)* 854–863.

[53] C. A. Cardona, J. A. Quintero, O. J. Sánchez, Challenges in Fuel Ethanol Production, *International Review of Chemical Engineering 1 (2009)* 581-597.

[54] J. B. Muniz, Effect of static non-homogeneous magnetic fields on the alcoholic fermentation by *Saccharomyces cerevisiae,* MSc Thesis, Federal University of Pernambuco, Recife-PE-Brazil, 2002 (*in Portuguese*).

[55] M. A. Motta, J. B. F. Muniz, A. Schuler, M.da Motta, Static magnetic fields enhancement of *Saccharomyces cerevisiae* ethanolic fermentation, *Biotechnology Progress 20 (2004)* 393-396.

[56] H. J. Lopes-Dahabada, M. Sola-Penna, Urea increases tolerance of yeast inorganic pyrophosphatase activity to ethanol: The other side of urea interaction with proteins. *Arch. Biochem. Biophys. 394 (2001)* 61-66.

[57] T. A. Galonja-Coghill, L. M. Kostadinovic, N. C. Bojat, Magnetically Altered Ethanol Fermentation Capacity of *Saccharomyces Cerevisiae*, *Proc. Nat. Sci, Matica Srpska Novi Sad, 117 (2009)* 119-123. Doi:10.2298/Zmspn0917119g.

[58] L. O. Santos, R. Monte Alegre, C. Garcia-Diego, J. Cuellar, Effects of magnetic fields on biomass and glutathione production by the yeast *Saccharomyces cerevisiae*, *Process Biochemistry 45 (2010)* 1362–1367 .

[59] J. E. M. Zapata, M. R. Hoyos, L. A. B. Quinchía, Kinetic Parameters of Growth of *Saccharomyces Cerevisiae* Affected by a Varying Magnetic Field of Low Intensity and High Frequency, *Vitae, Revista De La Facultad De Química Farmacéutica (Medelin, Colombia) 12 (2005)* 39-44 (in Spanish)

[60] J. E. Zapata, G. Moreno, E. J. Marquez, Magnetic field effects on the growth of *Saccharomyces cerevisiae*. *Interciencia 27 (2002)* 1-7.

[61] M. Ikehata, M. Iwasaka, J. Miyakoshi, S. Ueno,T. Koana, Effects of intense magnetic fields on sedimentation pattern and gene expression profile in budding yeast, *Journal of Appl Physics 93 (2003)* 6724–6726.

[62] V. H. Perez, A. F. Reyes, O. R. Justo, D. C. Alvarez, R. M. Alegre, Bioreactor Coupled with Electromagnetic Field Generator: Effects of extremely low frequency electromagnetic fields on ethanol production by *Saccharomyces cerevisiae* ,

*Biotechnology Progress 23 (2007)* 1091-1094. doi: 10.1021/bp070078k .

[63] L. Zhang, H. Berg, Electrostimulation of the dehydrogenase system of yeast by alternating currents, *Bioelectrochemistry and Bioenergetics 28 (1992)* 341-353.

[64] S. Nakasono, C. Laramee, H. Saiki, K. J. McLeod, Effect of power-frequency magnetic fields on genome-scale gene expression in *Saccharomyces cerevisiae*, *Radiation Research 160 (2003)* 25–37.

[65] D. M. Binninger, V. Ungvichian, Effects of 60 Hz AC magnetic fields on gene expression following exposure over multiple cell generations using *Saccharomyces cerevisiae*, *Bioelectrochemistry and Bioenergetics 43 (1997)* 83-89 .

[66] C. T .Rogers, P. J. Hore, Chemical magnetoreception in birds: The radical pair mechanism. *Proce. of the Natl. Acad. Sci., USA , 06 (2009)* 353–360.

[67] A. R. Liboff, *Biological effects of magnetics fields*, vol. 2:171-175 (Plenum Press, New York-London, 1969).

[68] S. Lindquist, Heat-shock proteins and stress tolerance in microorganisms, *Curr. Opin. Genet. Dev. 2 (1992)* 748–755.

[69] Y. Ye, Y. Zhu, L. Pan, L. Li, X. Wang, Y. Lin, Gaining insight into the response logic of *Saccharomyces cerevisiae* to heat shock by combining expression profiles with metabolic pathways, *Biochemical and Biophysical Research Communications 385 (2009)* 357–362. doi:10.1016/j.bbrc.2009.05.071.

[70] J. P. Hendrick, F. U. Hartl, Molecular chaperone functions of heat-shock proteins, *Annu. Rev. Biochem. 62 (1993)* 349–384.

[71] A. P. Gasch, P. T. Spellman, C. M. Kao, O. Carmel-Harel, M. B. Eisen, G. Storz, D. Botstein, P. O. Brown, Genomic expression programs in the response of yeast cells to environmental changes, *Mol. Biol. Cell 11 (2000)* 4241–4257.

[72] B. Panaretou, C. Zhai , The heat shock proteins: Their roles as multi-component machines for protein folding , *Fungal biology reviews 22 (2008)* 110–119. doi:10.1016/j.fbr.2009.04.002.

[73] D. A. Parsell, A. S. Kowal, S. Lindquist, *Saccharomyces cerevisiae* Hsp104 protein. Purification and characterization of ATP induced structural changes, *J. Biol. Chem. 269 (1994)* 4480–4487.

[74] Y. Sanchez, J. Taulien, K. A. Borkovich, S. Lindquist, Hsp104 is required for tolerance to many forms of stress. EMBO J. 11 *(1992)* 2357–2364.

[75] J. S. Hahn, Z. Hu, D. J. Thiele, V. R. Iyer, Genome-wide analysis of the biology of stress responses through heat shock transcription factor, *Mol. Cell. Biol. 24 (2004)* 5249–5256.

[76] T. Hottiger, T. Boller, A. Wiemken, Rapid changes of heat and desiccation tolerance correlated with changes of trehalose content in *Saccharomyces cerevisiae* cells subjected to temperature shifts, *FEBS Lett. 220 (1987)* 113–115.

[77] P. K. Sorger, H. R. B. Pelham, Yeast heat shock factor is an ssential DNA-binding protein that exhibits temperature-dependent phosphorylation, *Cell 54 (1988)* 855–864.

[78] T. Hottiger, C. De Virgilio, M. N. Hall, T. Boller, A. Wiemken, The role of trehalose synthesis for the acquisition of thermotolerance in yeast. II. Physiological concentrations of trehalose increase the thermal stability of proteins in vitro, *Eur. J. Biochem. 219 (1994)* 187–193.

[79] L. K. Conlin, H. C .M. Nelson, The Natural osmolyte trehalose is a positive regulator of the heat-induced activity of yeast heat shock transcription factor, *Mol. Cell. Biol. 27 (2007)* 1505–1515.

[80] J. F. Davidson, B. Whyte, P .H. Bissinger, R. H. Schiestl, Oxidative stress is involved in heat-induced cell death in Saccharomyces cerevisiae, *Proc. Natl. Acad. Sci. USA 93 (1996)* 5116–5121.

[81] G. Marchler, C. Schuller, G. Adam, H. Ruis, A Saccharomyces cerevisiae UAS element controlled by protein kinase A activates transcription in response to a variety of stress conditions, *EMBO J. 12 (1993)* 1997–2003.

[82] P. Zancan, M. Sola-Penna, Trehalose and glycerol stabilize and renature yeast inorganic pyrophosphatase inactivated by very high temperatures, *Arch. Biochem. Biophys. 444 (2005)* 52–60.

[83] L. Li, Y. R. Ye, L. Pan, Y. Zhu, S. P. Zheng, Y. Lin , The induction of trehalose and glycerol in *Saccharomyces cerevisiae* in response to various stresses , *Biochemical and Biophysical Research Communications 387 (2009)* 778–783. doi:10.1016/j.bbrc.2009.07.113.

[84] A. Blomberg, Metabolic surprises in *Saccharomyces cerevisiae* during adaptation to saline conditions: questions, some answers and a model, *FEMS Microbiol. Lett.* 182 *(2000)* 1–8.

[85] C. Ferreira, C. Lucas, Glucose repression over Saccharomyces cerevisiae glycerol/H$^+$ symporter gene STL1 is overcome by high temperature, *FEBS Letters 581 (2007)* 1923–1927. doi:10.1016/j.febslet.2007.03.086.

[86] R. J. Karreman, E. Dague, F. Gaboriaud , F. Quilès,J. F. L. Duval, G. G. Lindsey, The stress response protein Hsp12p increases the flexibility of the yeast , *Biochimica et Biophysica Acta 1774 (2007)* 131–137. doi:10.1016/j.bbapap.2006.10.009.

[87] M. Blank*,* R. Goodman, Electromagnetic fields stress living cells, *Pathophysiology* 16 *(2009)* 71–78.

[88] A. Pazur, C. Schimek, P. Galland, Magnetoreception in microorganisms and fungi, *Central European Journal of Biology, 2 (2007)* 597–659. doi: 10.2478/s11535-007-0032-z

## Authors' information

[1] Department of Chemical Engineering
University of Chemical Technology and Metallurgy (UCTM)
1756 Sofia, 8 Kl. Ohridsky Blvd., Bulgaria, e-mail: jordan.hristov@mail.bg ; hristovmeister@gmail.com ,
website : http://hristov.com/jordan

[2] Universidade Estadual do Norte Fluminense Darcy Ribeiro - UENF
Centro de Ciências e Tecnologias Agropecuárias – CCTA, Laboratório de Tecnologia de Alimentos – LTA, Av. Alberto Lamego, 2000. Prédio P4, Parque Califórnia , Campos dos Goytacazes – RJ , CEP 28013-602, Brazil , e-mail: victorh@uenf.br